\documentstyle[astrobib,epsfig]{mn}

\begin{document}

\def\bi#1{\hbox{\boldmath{$#1$}}}

\newcommand{\beq}{\begin{equation}}
\newcommand{\eeq}{\end{equation}}
\newcommand{\beqa}{\begin{eqnarray}}
\newcommand{\eeqa}{\end{eqnarray}}

\newcommand{\lexp}{\mathop{\langle}}
\newcommand{\rexp}{\mathop{\rangle}}
\newcommand{\rexpc}{\mathop{\rangle_c}}

\def\bi#1{\hbox{\boldmath{$#1$}}}

\title[Cluster abundance normalization from observed mass-temperature relation]{Cluster abundance normalization from observed mass-temperature relation}

\author[U. Seljak]{
 U.~Seljak\thanks{E-mail: uros@feynman.princeton.edu} \\
 Department of Physics, Princeton University, Princeton, NJ 08544,
 USA}

\pubyear{2001}

\maketitle

\begin{abstract}
Abundance of rich clusters in local universe 
is currently believed to provide the most robust 
normalization of power spectrum at a scale of 10 Mpc.
This normalization depends very sensitively on the 
calibration between virial mass $M$ and temperature $T$, which is usually 
taken from simulations. Uncertainties in the modelling, such 
as gas cooling and heating, can lead to a factor 
of two variations in the normalization and are thus not very reliable. 
Here we use instead 
an empirical $M_{500}-T$ relation derived from X-ray mass determinations to 
calibrate the method. 
%Here $M_{500}$ is the cluster mass at radius within 
%which the mean density is 500 times the critical density. 
We use results from dark matter simulations
to relate the virial mass function to the mass function at observed $M_{500}$.
We find that the best fitted value 
in flat models is 
$\sigma_8=(0.7\pm 0.06) (\Omega_m/0.35)^{-0.44}(\Gamma/0.2)^{0.08}$, 
where only the statistical error is quoted.
This is significantly 
lower than previously obtained values from the local cluster abundance.
This lower value for $\sigma_8$ is in a better agreement
with cosmic microwave background and large scale structure
constraints and helps alleviate 
small scale problems of CDM models. 
Presently the systematic uncertainties in the mass determination
are still large, but
ultimately this method should provide a more 
reliable way to normalize the $M-T$ relation. This can 
be achieved by obtaining a larger sample of well measured 
cluster masses out to a significant fraction of 
virial radius with BeppoSAX, Chandra and XMM-Newton.

\end{abstract}

%\keywords{large-scale structure of universe}

\section{Introduction}

Abundance of clusters in the local universe provides one of the 
strongest constraints on the matter power spectrum amplitude 
on scales of order 10Mpc. The main power of this method is that 
the abundance depends very sensitively on the amplitude 
of {\it linear} power spectrum on this scale. 
Numerous studies over the past decade have investigated this 
constraint, obtaining more or less consistent results (\citeNP{1993MNRAS.262.1023W}, 
\citeNP{1996ApJS..103...63B}, \citeNP{1996MNRAS.281..323V}, 
\citeNP{1997ApJ...490..557K},
\shortciteNP{1998MNRAS.298.1145E}, \citeNP{1998ApJ...498...60P},
\citeNP{1998ApJ...508..483W},
\citeNP{2000A&A...362..809B}, \citeNP{2000ApJ...534..565H}). Two 
most recent studies are those by \citeN{2001MNRAS.325...77P}
using a local sample of 
clusters and \shortciteN{2001ApJ...561...13B} using a higher redshift sample, both of 
which find a relative error of $5-10\%$ on the density dependent
normalization (but which are inconsistent with each other at a more than
3-$\sigma$ level, see discussion below). 

While cluster abundance normalization is usually assumed to be the most 
reliable on a scale of 10Mpc, other data sets also provide some 
more model dependent constraints on the same scale. 
Recent determinations of the best fitted model from CMB combined with 
large scale structure
(\citeNP{2001astro.ph..5091W}, \shortciteNP{2001astro.ph..9152E}) and from Ly-$\alpha$ forest (\citeNP{1999ApJ...520....1C}, \citeNP{2000ApJ...543....1M}) both give $\sigma_8$ 
values more than 30\% lower than those from the latest 
local cluster abundance determinations for the 
favored $\Omega_m=0.3-0.4$ case.
A lower normalization also helps significantly with the problems
that CDM has on small scales \cite{2001astro.ph..9392K}. This tension between 
the cluster abundance normalization and other data sets, which is 
formally at more than 3-$\sigma$ level, could indicate 
there are still some unrecognized 
systematic uncertainties present in the cluster abundance normalization. 

An implicit assumption in all of the studies above
is to assume a relation 
between halo mass $M$ and observed gas temperature $T$, 
$M/(10^{15}h^{-1}M_{\sun})=(T/\beta)^{3/2}\Delta_c^{-1/2}E^{-1}$,
where $\Delta_c$ is the mean overdensity inside
the virial radius defined using spherical collapse model, 
$T$ is gas temperature in keV, 
$E^2=\Omega_m(1+z)^3+\Omega_{\Lambda}$ for the
zero-curvature case and
the constant $\beta$ is determined 
from the hydrodynamical simulations. The problem with this approach
is that the simulations may not capture all of the physics occuring 
in the real world. 
For example, most simulations only include adiabatic hydrodynamics, while
it was recently shown 
that including gas cooling 
(which must be present at some level)
can lead to a factor of two change in the normalization 
constant of the mass temperature relation 
\shortcite{2001ApJ...552L..27M}. Other 
potentially important effects, such as heating of the gas with stars, 
supernovae or AGN, 
could further change the relation.  
Resolution issues could also be affecting the results: 
there is quite a lot of 
scatter in the relation between different groups even when all the 
physical ingredients are 
identical %\footnote{Some of this variation could also be due to the fact that 
%different groups simulate different cosmological models. The
%constant of proportionality $\beta$ 
%could vary with cosmology, since $\Delta_c$ is based 
%on spherical collapse 
%model, which may have little relation to reality.} 
(see \citeNP{2001MNRAS.327.1353K} and \shortciteNP{2001MNRAS.325...77P} for a 
recent comparison between different groups).
Yet another possible complication is the observer's
definition of temperature, which depends on the spectral bandpass
and other instrumental details. 
\citeN{2001ApJ...546..100M} explore several different possibilities and 
find up to a 20\% variation between them. 
Since all these effects are systematic it is difficult to place
a reliable error estimate on them, except to argue that a factor of 
two changes are not excluded at this point. To put this into a 
context, such a change leads to a 30-40\% change in the overall power 
spectrum normalization, much larger than the quoted error-bars.
It seems therefore worthwhile to explore alternative methods to 
normalize the $M-T$ relation which do not rely exclusively 
on the simulations.

Over the past several years
there has been a lot of progress in the direct determination 
of cluster masses using spatially resolved temperature and X-ray
intensity information. Under the assumption of spherical symmetry and 
hydrostatic equilibrium with no non-thermal pressure support 
one can solve for the radial mass profile of  
the matter in the cluster. This approach 
consistently yields $M-T$ relation up to a factor of two lower than the average 
over simulations (\citeNP{1999ApJ...520...78H}, \citeNP{2000ApJ...532..694N}, 
\citeNP{2001A&A...368..749F}, \citeNP{2001astro.ph.10610A}). 
Such a discrepancy could be 
explained if there was an additional non-thermal pressure 
support, which should be added to the equation of hydrostatic
equlibrium (such as from turbulent motions or magnetic fields). 
However, recent comparison between X-ray mass measurements and weak 
lensing mass measurements for several clusters 
shows very good agreement between the 
two, excluding the possibility of a large non-thermal support, at least 
for relaxed clusters
\shortcite{2001astro.ph.10610A}. An alternative possibility 
is to assume that X-ray mass 
determinations are reliable 
and there is some systematic problem 
with the simulations. 

The purpose of this paper is to use the empirical $M-T$ relation 
to derive the power spectrum normalization. An earlier 
attempt to do this based on
one measured cluster mass has been made 
by \citeN{1998ApJ...504...27M} and a lower value of $\sigma_8$ has been 
found than calibrating from simulations. Since then the observed 
$M-T$ relation and scatter around it has 
been established much more accurately.
Since the measured masses are limited to the inner parts of 
the cluster they cannot be extended to the virial
mass,
defined here as the mass that gives rise to the universal mass function, 
without additional modelling.  
Here we use CDM type 
mass profiles to make this connection.
We find that 
the offset relative to 
the simulations persists and leads to a significantly lower
power spectrum normalization than was found before.
The implications of this result and future prospects for this approach 
are discussed in the conclusions.

\section{Theory}

The halo mass function describes the number density of halos
as a function of mass. It can be written as
\begin{equation}
{dn \over d\ln M} ={\bar{\rho} \over M}f(\sigma){d \ln \sigma^{-1} \over 
d \ln M},
\end{equation}
where $\bar{\rho}$ is the mean matter density of the universe, $M$ is the 
virial mass of the halo and $n(M)$ is the spatial number density of halos 
of a given mass $M$. We
introduced
function $f(\sigma)$, which 
has a universal form independent of the
power spectrum, matter density, normalization or redshift if written
as a function of rms variance of linear density field $\sigma$, 
\begin{equation}
\sigma^2(M)=4\pi \int P(k)W_R(k)k^2 dk.
\end{equation}
Here $W_R(k)$ is the Fourier transform
of the spherical top hat window with radius $R$, chosen such that it encloses 
the mass $M=4\pi R^3 \bar{\rho}/3$ and $P(k)$ is the linear power spectrum.

The universality of the
mass function has been recently investigated by a number 
of authors (\citeNP{1999MNRAS.308..119S}, 
\shortciteNP{2001MNRAS.321..372J}, \citeNP{2001A&A...367...27W}),
where it has been shown that
the mass function is indeed universal for a broad range of cosmological 
models. \shortciteN{2001MNRAS.321..372J}
propose the form
\begin{equation}
f(\sigma)=0.315 \exp[-|\ln \sigma^{-1}+0.61|^{3.8}].
\end{equation}
They find this equation works best if the cluster mass $M$ is defined as 
the mass within the radius where mean density in units of 
critical
is 200$\Omega_m$. Other authors (\citeNP{1999MNRAS.308..119S}, 
\citeNP{2001A&A...367...27W}) 
find that a similar mass function
works if the mass is defined within the spherical overdensity defined 
by the spherical collapse model, which is 108 for $\Lambda$CDM model 
with $\Omega_m=0.35$ that we adopt here as the reference model. 
We will adopt the \shortciteN{2001MNRAS.321..372J} 
value here and comment on the implications of other choice
in the discussion section. 

\begin{figure}
  \begin{center}
    \leavevmode\epsfxsize=8cm \epsfbox{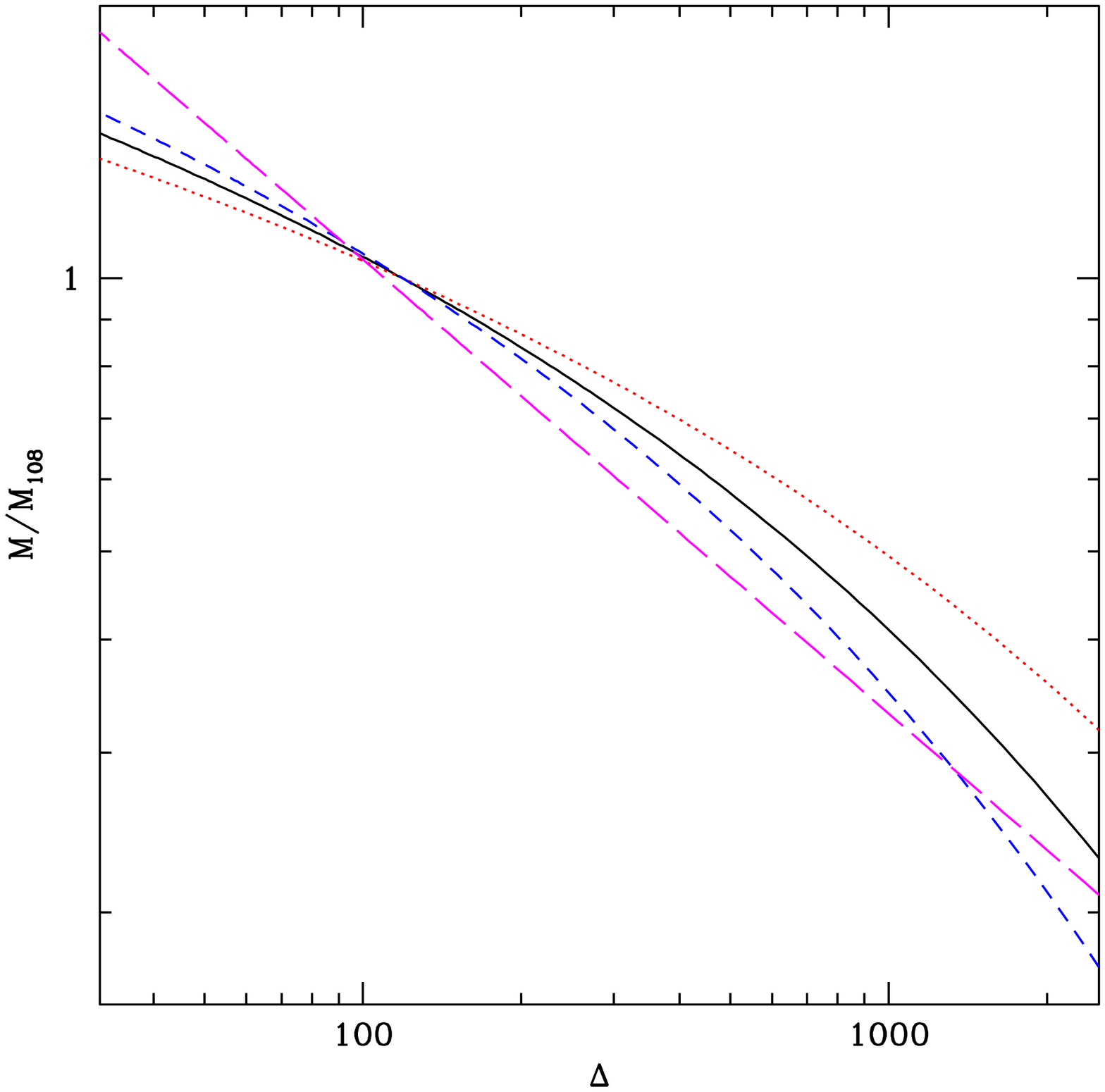}
%\leavevmode
%\epsfxsize=8cm \epsfbox{fig1.ps}
%\epsfig{file=fig1.ps, width=8cm}
\end{center}
\caption{$M_{\Delta}/M_{\Delta_c}$ versus $\Delta$ for $c_{\Delta_c}=3.7$
(short dashed), $c_{\Delta_c}=5$ (solid) and $c_{\Delta_c}=7.5$ (dotted)
for $\Delta_c=108$. Also shown is isothermal profile $M_{\Delta}/M_{\Delta_c}
= (\Delta/\Delta_c)^{-1/2}$ as long dashed line. 
}
\label{fig1}
\end{figure}

Given that the mass profiles at large radii are difficult to measure
observers prefer to quote the masses $M_{\Delta}$ at smaller radii
or higher $\Delta$ than virial values.
Two often used values in the literature are $M_{2500}$ 
\shortcite{2001astro.ph.10610A} 
and $M_{500}$ (e.g. \shortciteNP{2001A&A...368..749F}). Since at the moment 
we do not have 
mass functions at $M_{500}$ or $M_{2500}$ directly from simulations 
we will construct them by relating $M_{\Delta}$
to $M_{\rm vir}$
adopting an average mass profile for the clusters. This is typically 
parametrized as \cite{1997ApJ...490..493N}
\begin{equation}
\rho(r)={\rho_s \over (r/r_s)^{-\alpha}(1+r/r_s)^{3+\alpha}}.
\label{rho}
\end{equation}
This model assumes that the profile shape is
universal in units of scale radius $r_s$, while its characteristic density
$\rho_s$ at $r_s$ or concentration $c_{\Delta_c}=r_{\Delta_c}/r_s$ 
may depend on the halo mass.
The halo profile is assumed to scale as
$r^{-3}$ in the outer parts and as $r^{\alpha}$
in the inner parts, with the transition between the two at $r_s$.
We fix the inner slope to
$\alpha=-1$ \shortcite{1997ApJ...490..493N}, 
since we are not concerned about the shape of the halo
profile in the center. 
The outer slope $r^{-3}$ is the most common value found in
the $N$-body simulations, although scatter around this value can be
quite significant \cite{2001MNRAS.324..450T}.
Even if we adopt this value the profile is still not fixed and the 
remaining freedom can be parametrized with the concentration parameter 
$c_{\Delta_c}$. 
In figure \ref{fig1} we show mass as a function of spherical overdensity 
in units of $M_{\Delta_c}$ for $c_{\Delta_c}=3.7, 5, 7.5$, which spans 
the range of concentration parameters appropriate for massive cluster halos.
Also shown is the isothermal profile sometimes used to describe the 
cluster. %The overdensity used here is $\Delta_c=108$ 
%based on spherical collapse model with $\Omega_m=0.35$. 
In the following 
we adopt value $c_{\Delta_c}=5$, which is consistent with both observational 
constraints \cite{2001astro.ph.10610A} and theoretical predictions 
(\shortciteNP{1997ApJ...490..493N}, \shortciteNP{2001MNRAS.321..559B}). 
We will ignore the small expected variation of $c_{\Delta_c}$ with cluster 
mass.  
Based on figure \ref{fig1} we estimate 
the error induced by this choice to be of order of 10\% if $M_{500}$ 
is used to determine $M_{200\Omega_m}$. 
This error rises to 20-30\% if  $M_{2500}$ is used instead
to determine $M_{200\Omega_m}$. %The error is slightly larger 
%if $\Delta_c=200 \Omega_m$ is used instead of the spherical collapse 
%model, since this extrapolates the profile to a somewhat larger radius. 

\section{Results}

The observed $M-T$ relation for $T>3{\rm keV}$ clusters relevant for 
abundance studies as given by \shortciteN{2001A&A...368..749F} is 
\begin{equation}
M_{200\Omega_m} =10^{15}h^{-1}(1\pm 0.2) 
M_{\sun}\left( {T \over \beta\Delta_c^{1/3}{\rm keV}}\right)^{1.5}, 
\label{mt}
\end{equation}
where $\beta = 1.75 \pm 0.25$ and 
we used the relation between $M_{500}$ and $M_{200\Omega_m}$ based on 
the adopted cluster profile (figure \ref{fig1}). 
The value of $\beta$ quoted above 
is for the fiducial model with $\Omega_m=0.35$, 
but in 
fact it varies little with $\Omega_m$, increasing only by 5\% for $\Omega_m=1$.
This is because while $\Delta_c$ in equation \ref{mt} increases with $\Omega_m$ 
the mass $M_{200\Omega_m}$ extrapolated 
from $M_{500}$ decreases, compensating each other for a profile that is 
close to isothermal. 

While the scaling in equation \ref{mt} 
is in agreement with the theoretical predictions the 
normalization of the $M-T$ relation is not. 
If we used $M_{\Delta_c}$ as done in previous work we would 
find $\beta = 1.95 \pm 0.25$.
This value
should be compared to $\beta=1.3\pm 0.2$ as the average 
over the simulations \shortcite{2001MNRAS.325...77P}. 
The two values are inconsistent with 
each other at a 3-$\sigma$ level, suggesting that there is a significant 
systematic error either in the observed mass estimates 
or in the simulations.
Other data samples find similar 
discrepancies between simulated and observed relation. For example, 
\shortciteN{2001astro.ph.10610A} find a 40\% offset in $M_{2500}-T_{2500}$ 
relation relative to the simulations. 
It is unclear how $T_{2500}$ obtained from 
Chandra relates to temperatures in cluster samples used to derive the
power spectrum normalization, which are mostly obtained from ASCA. 
Moreover, $M_{2500}$ to $M_{\Delta_c}$ 
conversion
carries considerable uncertainty (figure \ref{fig1}). 
For these reasons we will not use
this data set here. %, although as the data set grows and temperature 
%calibrations become available this should 
%become one of the best ways to reduce the current uncertainties in the 
%normalization. With better data one could use
%the concentration parameter values obtained from 
%fits to the actual profiles \shortcite{2001astro.ph.10610A}, 
%further reducing the uncertainty in virial mass determination. 

For the cluster sample we use a recent
compilation by \shortciteN{2001MNRAS.325...77P} 
based on the data from nearby clusters. Their 
adopted temperature measurements are taken mostly from 
\citeN{2000MNRAS.312..663W}, which are
on average 15\% higher than those derived by \citeN{1998ApJ...504...27M}. 
This by itself
can be a significant systematic effect. Since we wish to compare this 
data sample to the $M-T$ relation by \shortciteN{2001A&A...368..749F} 
we must compare the 
temperatures of the two samples. We do this in two different ways. 
First, we compare the derived 
luminosity-temperature ($L-T$) relation for the two samples, which 
gives us the relative 
normalization between the adopted temperature values, since 
in both samples luminosities are obtained from ROSAT. 
\shortciteN{2001A&A...368..749F} data combined with $L-M$ relation \cite{1999dtrp.conf..157R} gives 
$L=1.15\times 10^{44}(T/6{\rm keV})^2h^{-2}{\rm ergs/s}$, which implies that temperature
values used by \shortciteN{2001MNRAS.325...77P} 
are on average 5\% higher than those used
by \shortciteN{2001A&A...368..749F}. Second, 13 clusters are contained in both samples. Linear 
regression with zero offset gives again the same 5\% deviation. 
The two methods give consistent results and 
we apply this small correction in the analysis below. 
Another small effect which goes in the same direction 
is to correct back the redshift dependence of $T$ which was 
applied by \shortciteN{2001A&A...368..749F} in the $M-T$ fitting. Together these 
corrections make an effective $\beta=1.9$ for the fiducial 
model. 

We emphasize that $T$ is used here just as a label and does not necessarily 
need to be the correctly calibrated temperature of the cluster, 
as long as the mass assigned to this $T$ is correct and there is a 
monotonic relation between the two. 
For this method to be unbiased we must only address the bias in the 
mass determination. However,
the masses 
are derived from the assumption of hydrostatic equlibrium combined with the
measurements of the temperature and gas profile. For a polytropic profile $T\propto 
\rho_{\rm gas}^{\gamma-1}$ and beta-model for X-ray intensity 
profile $I_{\rm X}
\propto  \left[1+\left(\frac{r}{r_{\rm c}}\right)^2\right]^{1/2-3\beta_{\rm X}}$ 
one finds 
\begin{equation}
M(r)\propto T\beta_{\rm X}\gamma{r^3\over r_{\rm c}^2+r^2}.
\end{equation}
Here $r_c$ is the core radius of the beta-model profile. 
We see that the derived mass is linear in $\beta_{\rm X}$, 
$\gamma$ and $T$ (note 
that the $\beta_{\rm X}$ here should not be confused with $\beta$ as a parametrization 
of $M-T$ relation), each of which has some uncertainty associated with it. 
Uncertainty in mass determination from 
surface brightness profile fitting is expected to be 
small. Somewhat
more controversial are $T$ profiles derived from ASCA, which have been 
interpreted by some groups (e.g. \shortciteNP{2001A&A...368..749F}, 
\citeNP{1998ApJ...504...27M})
to show a decline in $T$ at larger radii, in agreement with 
theoretical predictions,% (see e.g. \cite{2001MNRAS.327.1353K}), 
while others do not find this (e.g. \citeNP{2000MNRAS.312..663W}). 
Such declines are not seen in general 
from Chandra \shortcite{2001astro.ph.10610A}, 
although the radii probed are  
smaller and this is not unexpected. 
BeppoSAX does see a decline in most of the clusters observed \cite{2001astro.ph.10469D} and 
for the few clusters that overlap with the 
ASCA sample  in \shortciteN{2001A&A...368..749F}
the drop in $T$ out to $r_{500}$
is in a reasonable agreement between the two. It is also in 
agreement with XMM-Newton for the clusters that have been observed by both 
satellites, 
although 
the XMM-Newton sample is still small \shortcite{2001A&A...365L..80A}.
The measured average polytropic index 
$\gamma \sim 1.2$ \shortcite{2001A&A...368..749F} is close 
to theoretical predictions \cite{2001MNRAS.327.1353K} and we verified 
that this 
also provides a reasonable fit to the BeppoSAX data \cite{2001astro.ph.10469D}
in the outer parts of the cluster where most of the mass is (in the inner 
parts the polytropic assumption fails, since it cannot predict a 
sharp flattening 
of the temperature profile at a radius larger than the core radius, as
seen with BeppoSAX data).

To derive the constraints on the power spectrum amplitude we compute the 
cumulative $T$ function at an effective mean redshift of the 
sample $z\sim 0.05$. We compare it to match the best fitted model of 
\shortciteN{2001MNRAS.325...77P} 
around $T=6.5$keV, which is the pivot point where most of 
the sensitivity is. The results are not very sensitive to this 
choice since the shapes of cumulative mass function are 
very similar. Our simplified approach, which avoids a proper 
likelihood evalution, is reasonable as long as the estimated errors 
in the two methods are comparable. Large errors can introduce bias 
since for a steeply declining $T$ function low $T$ clusters can 
scatter into high $T$ and produce a cumulative 
$T$ function with a higher amplitude in this exponential regime 
\cite{1998ApJ...498...60P}. 
The scatter in $M-T$ relation is comparable or perhaps somewhat larger 
than the estimated errors in \shortciteN{2001MNRAS.325...77P}.
In the latter case this would lead to a slight overestimate of $\sigma_8$.
Here we will assume the errors are comparable and we do not correct for
this effect, which would further reduce $\sigma_8$. 

\begin{figure}
  \begin{center}
    \leavevmode\epsfxsize=8cm \epsfbox{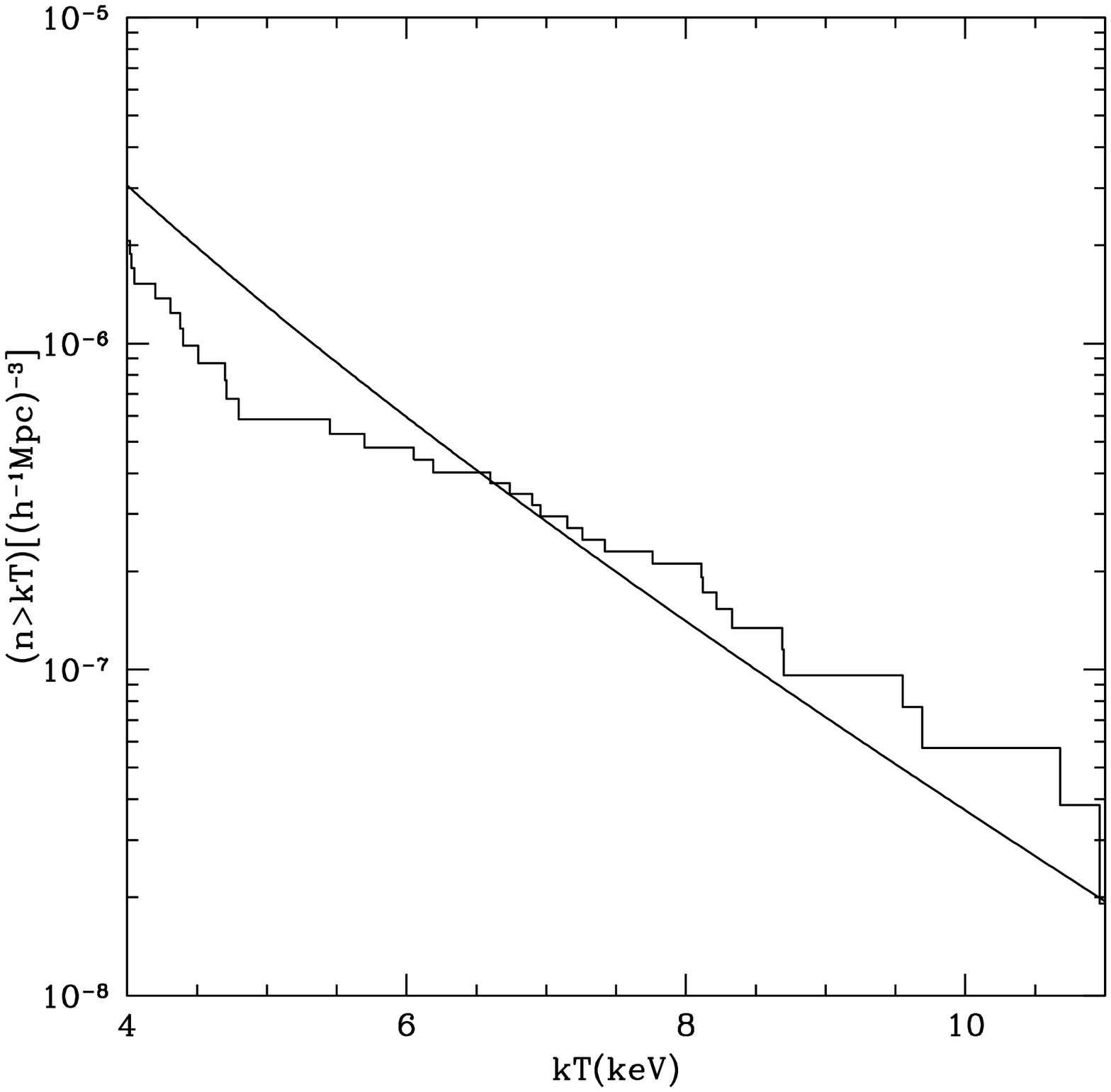}
%\leavevmode
%\epsfxsize=8cm \epsfbox{fig2.ps}
%\epsfig{file=fig1.ps, width=8cm}
\end{center}
\caption{
Cumulative temperature function constructed by 
\protect\shortciteN{2001MNRAS.325...77P} compared to the best fitted model 
for a flat cosmology with $\Omega_m=0.35$.
}
\label{fig2}
\end{figure}

Figure \ref{fig2} shows the comparison between the best fitted theoretical 
prediction for $\Omega_m=0.35$, $\Omega_{\Lambda}=0.65$, $\Gamma=0.2$ model
and the data sample of \shortciteN{2001MNRAS.325...77P}. Here $\Gamma$ 
is the shape parameter of the power spectrum. The best fitted 
value for this model is $\sigma_8=0.7$. 
A more general fit around this model is 
$\sigma_8=0.7(\Omega_m/0.35)^{-0.44}(\Gamma/0.2)^{0.08}$, which 
is supposed to give a few percent accuracy over $0.2<\Omega_m<1$ and 
$0.1<\Gamma<0.3$. Note that the dependence on $\Omega_m$ is less steep than 
the usual $\Omega_m^{-0.6}$, a consequence of the different mass temperature 
relation and different mass function and mass definition. 
We estimate the error by varying $\beta$ by one 
standard variation in both directions. This gives $\sigma_8=0.7 \pm 0.06$ 
for the fiducial model. We emphasize that this is only the statistical 
error. The systematical error, which is much harder to estimate, is 
discussed below.

\section{Discussion and conclusions}\label{conclusions}

The value for $\sigma_8 \sim 0.7\pm 0.06$ for $\Omega_m=0.35$ is about 25\% 
lower than the most recent value derived from the local cluster sample 
\shortcite{2001MNRAS.325...77P}. 
While it is very close to the value obtained from high 
redshift cluster catalog \cite{2001ApJ...561...13B}, it is important to note 
that they use $\beta=1.2$
in their analysis and so if the actual value is $\beta\sim 2$ as 
suggested by the local $M_{500}-T$ 
observations then their value for $\sigma_8$ would 
be even lower. For the same value of $\beta$ 
the discrepancy in the normalization 
between the local and the 
high redshift sample is at a more than 3-$\sigma$ level and 
difficult to explain as a statistical fluctuation. This points to perhaps
additional systematic effects beyond the $M-T$ normalization 
discussed here that may be affecting either the local or the 
high redshift sample. We also note that
our lower value for $\sigma_8$ agrees well with other recent attempts to 
circumvent the use of simulations by adopting the measured masses 
from weak lensing \cite{2001astro.ph.11394V} 
and hydrostatic equilibrium \cite{2001astro.ph.11285R}. 
The latter analysis in particular is very close in spirit to our approach.

Even though the method used here does not rely on simulations
and so in principle should be
more reliable, it still contains several 
possible systematic uncertainties. We mentioned some before: 
the mass determination from spatially resolved temperature maps has 
perhaps a 20\% uncertainty. 
It is difficult, although perhaps not impossible, to invoke systematic 
errors in the mass determination to bring 
the observed $M-T$ relation  
in agreement with the adiabatic simulations results.
We know however that additional processes must be included 
in simulations. Cooling, which is relatively well understood, 
although numerically challenging, makes a significant difference in 
the normalization of $M-T$ relation and its inclusion 
brings the predicted relation closer to the observations 
\cite{2001ApJ...552L..27M}.
Underestimation of
errors in $M-T$ can also bias the results 
and if these are larger than assumed here 
then $\sigma_8$ will be reduced further, although probably by not more 
than a few percent unless the scatter in $M-T$ relation is seriously 
underestimated.

On the theoretical side the 
extrapolation from $M_{500}$ to $M_{\Delta_c}$ has roughly a 10\% uncertainty 
in mass. 
The use of 
the mass function by \citeN{1999MNRAS.308..119S} with $M_{\Delta_c}$  
rather than \shortciteN{2001MNRAS.321..372J} mass function with $M_{200 \Omega_m}$
would decrease the value of $\sigma_8$ by 7\%. 
Rather than to construct 
mass function based on  
a spherical overdensity at a virial radius, 
which is not directly observable, it 
is more useful to focus on mass functions using masses defined within some 
smaller radius, such as $r_{500}$, where mass can be directly 
measured. A step in this direction has been undertaken recently by 
\shortciteN{2001astro.ph.10246E}, 
who evaluate mass function at $M_{200}$, which is only 40\%
above $M_{500}$ for $c_{\Delta_c}=5$. 
Unfortunately they do not provide a universal
mass function valid for all models, but a fitting formula for the two
cosmological models they simulate. 
We find that for flat $\Omega_m=0.3$ model their mass function agrees with ours 
if the mass definition of \shortciteN{2001MNRAS.321..372J}
is used together with $c_{\Delta_c}=5$. Similarly, M. White (private
communication) has evaluated mass functions using $M_{500}$ for several
models and again we find the agreement with our mass function is very good.
This can be further improved in the future by defining the mass as close 
as possible to the observational definition (e.g. in projection for 
the case of lensing and X-rays). The agreement between different 
groups gives us confidence that the mass function has converged to the 
form proposed by \shortciteN{2001MNRAS.321..372J} and this
should not be a major source of uncertainty in the future. On the other hand, 
use of different mass functions can explain up to 10\% differences in 
derived $\sigma_8$ in the past work.

Our results are actually good news for CDM models, which are facing 
significant problems on small scales (see \shortciteNP{2001astro.ph..9392K}
for an overview). The models predict too steep 
density profiles of halos and too many subhalos within halos, 
both of which are 
related to predicting too much small scale power. If the 10Mpc 
normalization can be significantly reduced this not only reduces the 
amplitude of the power spectrum, but also allows for a stronger tilt, 
which can further reduce the power on scales below 1Mpc where most of 
the problems are. 
Such tilted $\Lambda$CDM model was recently investigated by \shortciteN{2001astro.ph..9392K}.
It was shown there that the halo structure
problems with CDM are significantly alleviated and the tilted low 
$\sigma_8$ model gives
halo concentrations in line with observations. 
Subhalo abundances are reduced as well, but not to the point 
where there would be too few to mach the observations once 
photoionization suppression is included \cite{2001astro.ph.11005B}.  

Perhaps the most important message of the present paper is 
that the systematic errors associated with the cluster abundance 
normalization are still significantly larger than often assumed
(see also \citeNP{2000ApJ...543..113V})
and so one should be cautious when ruling out cosmological 
models based on this constraint. 
Our results suggest that even models with $\sigma_8\sim 0.6$ 
at $\Omega_m=0.3$ are not excluded at this point.
Although at present the cluster normalization may be less reliable 
than previously thought, the prospects for the future are more 
promising.  Instead of concentrating on 
calibration of $M-T$ relation from the simulations, which may be
uncertain for some time in the future, one should focus 
on obtaining better data to empirically calibrate the relation 
through the mass determinations from 
lensing, velocity dispersions
and applications of hydrostatic equlibrium. The latter should be
especially promising as well calibrated spatially resolved temperature profiles 
become available with Chandra, BeppoSAX and XMM-Newton. These mass 
measurements should be tied to the temperature determinations of clusters 
used in the construction of cluster sample and compared to mass functions 
constructed using masses which are observationally defined. 
As larger samples 
become available one will be able to construct mass selected samples directly 
without the need of using temperatures at all.
Ultimately this approach
should provide a reliable 
constraint on the cosmological models at 10Mpc scale.

The author acknowledges support from NASA,
Packard Foundation and Sloan Foundation. I would like to thank 
Stefano Borgani, Elena Pierpaoli and Martin White for useful comments and 
discussions. I also thank Douglas Scott for 
providing the data for figure \ref{fig2}. 

     \bibliography{apjmnemonic,cosmo,cosmo_preprints}
	\bibliographystyle{mnras}

\end{document}